\documentclass[a4paper,aps,prl,twocolumn,superscriptaddress,nobibnotes,showpacs]{revtex4}
\usepackage[utf8x]{inputenc}
\usepackage[english]{babel}
\usepackage{fontenc}
\usepackage{graphicx}
\usepackage{color}
\usepackage{multirow}
\usepackage{todonotes}
\usepackage{amsmath,amssymb,amsbsy}

\begin{document}

\title{Precision mass measurements of $^{58-63}$Cr: nuclear collectivity towards the \emph{N}=40 island of inversion}

\author{M. Mougeot}
\altaffiliation{Corresponding author : maxime.mougeot@csnsm.in2p3.fr}
\affiliation{CSNSM-IN2P3-CNRS, Universit\'{e} Paris-Sud, 91406 Orsay, France}

\author{D. Atanasov}
\altaffiliation{Present Address : Institut f\"{u}r kern- und Teilchenphysik, Institut f\"{u}r kern- und Teilchenphysik, }
\affiliation{Max-Planck-Institut f\"{u}r Kernphysik, Saupfercheckweg 1, 69117 Heidelberg, Germany}

\author{K. Blaum}
\affiliation{Max-Planck-Institut f\"{u}r Kernphysik, Saupfercheckweg 1, 69117 Heidelberg, Germany}

\author{K. Chrysalidis}
\affiliation{CERN, 1211 Geneva, Switzerland}
\affiliation{Insitut f\"{u}r Physik, Johannes Gutenberg-Universit\"{a}t, D-55099 Mainz, Germany}

\author{T. Day Goodacre}
\altaffiliation{Present Address : TRIUMF, 4004 Wesbrook Mall, Vancouver BC V6T 2A3, Canada}
\affiliation{CERN, 1211 Geneva, Switzerland}
\affiliation{School of Physics and Astronomy, The University of Manchester, Manchester M13 9PL, United Kingdom }

\author{D. Fedorov}
\affiliation{Petersburg Nuclear Physics Institute, 188300, Gatchina, Russia}

\author{V. Fedosseev}
\affiliation{CERN, 1211 Geneva, Switzerland}

\author{S. George}
\altaffiliation{Present Address : Universit\"{a}t Greifswald, Institut f\"{u}r Physik, 17487 Greifswald, Germany}
\affiliation{Max-Planck-Institut f\"{u}r Kernphysik, Saupfercheckweg 1, 69117 Heidelberg, Germany}

\author{F. Herfurth}
\affiliation{GSI Helmholtzzentrum f\"{u}r Schwerionenforschung GmbH, 64291 Darmstadt, Germany}

\author{J.D. Holt}
\affiliation{TRIUMF, 4004 Wesbrook Mall, Vancouver BC V6T 2A3, Canada}
 
\author{D. Lunney}
\affiliation{CSNSM-IN2P3-CNRS, Universit\'{e} Paris-Sud, 91406 Orsay, France}

\author{V. Manea}
\affiliation{CERN, 1211 Geneva, Switzerland}

\author{B. Marsh}
\affiliation{CERN, 1211 Geneva, Switzerland}

\author{D. Neidherr}
\affiliation{GSI Helmholtzzentrum f\"{u}r Schwerionenforschung GmbH, 64291 Darmstadt, Germany}

\author{M. Rosenbusch}
\altaffiliation{Present Address : RIKEN Nishina Center for Accelerator-Based Science, Wako, Saitama 351-0198, Japan
}
\affiliation{Universit\"{a}t Greifswald, Institut f\"{u}r Physik, 17487 Greifswald, Germany}
 
\author{S. Rothe}
\affiliation{CERN, 1211 Geneva, Switzerland}
 
\author{L. Schweikhard}
 \affiliation{Universit\"{a}t Greifswald, Institut f\"{u}r Physik, 17487 Greifswald, Germany}
 
\author{A. Schwenk}
\affiliation{Institut f\"ur Kernphysik, Technische Universit\"at Darmstadt, 64289 Darmstadt, Germany}
\affiliation{ExtreMe Matter Institute EMMI, GSI Helmholtzzentrum f\"ur Schwerionenforschung GmbH, 64291 Darmstadt, Germany}
\affiliation{Max-Planck-Institut f\"{u}r Kernphysik, Saupfercheckweg 1, 69117 Heidelberg, Germany}

 \author{C. Seiffert}
\affiliation{CERN, 1211 Geneva, Switzerland}

\author{J. Simonis}
\affiliation{Institut f\"{u}r Kernphysik and PRISMA Cluster of Excellence, Johannes Gutenberg-Universit\"{a}t, 55099 Mainz, Germany }
\affiliation{Institut f\"ur Kernphysik, Technische Universit\"at Darmstadt, 64289 Darmstadt, Germany}
\affiliation{ExtreMe Matter Institute EMMI, GSI Helmholtzzentrum f\"ur Schwerionenforschung GmbH, 64291 Darmstadt, Germany}

\author{S.R. Stroberg}
\affiliation{TRIUMF, 4004 Wesbrook Mall, Vancouver BC V6T 2A3, Canada}

\author{A. Welker}
\altaffiliation{Present Address : CERN, 1211 Geneva, Switzerland}
\affiliation{Institut f\"{u}r kern- und Teilchenphysik, Technische Universit\"{a}t Dresden, 01069 Dresden, Germany}

\author{F. Wienholtz}
\altaffiliation{Present Address : CERN, 1211 Geneva, Switzerland}
\affiliation{Universit\"{a}t Greifswald, Institut f\"{u}r Physik, 17487 Greifswald, Germany}
 
\author{R. N. Wolf}
\altaffiliation{Present Address : ARC Centre of Excellence for Engineered Quantum Systems, The University of Sydney, NSW 2006, Australia}
\affiliation{Max-Planck-Institut f\"{u}r Kernphysik, Saupfercheckweg 1, 69117 Heidelberg, Germany}

\author{K. Zuber}
\affiliation{Institut f\"{u}r kern- und Teilchenphysik, Technische Universit\"{a}t Dresden, 01069 Dresden, Germany}

\date{October 19, 2017}

\begin{abstract}
The neutron-rich isotopes $^{58-63}$Cr were produced for the first time at the ISOLDE facility and their masses were measured with the ISOLTRAP spectrometer. The new values are up to 300 times more precise than those in the literature and indicate significantly different nuclear structure from the new mass-surface trend. A gradual onset of deformation is found in this proton and neutron mid-shell region, which is a gateway to the second island of inversion around \emph{N}=40. In addition to comparisons with density-functional theory and large-scale shell-model calculations, we present predictions from the valence-space formulation of the \emph{ab initio} in-medium similarity renormalization group, the first such results for open-shell chromium isotopes.
\end{abstract}


\maketitle

Over the last few decades, the stability of proton and neutron ``magic'' numbers has been a major focus of experimental nuclear physics. Early momentum came from the vanishing of the \emph{N}=20 shell closure near $^{32}$Mg in mass measurements \cite{PhysRevC.12.644} from which arose the idea of the ``island of inversion''.

Extensive effort has followed to examine the classical signatures for magicity in exotic nuclei (e.g empirical shell gap or the energy of the first excited 2$^{+}_{1}$ state). More than three decades later, the robustness of all major shell closures has been assessed \cite{Sorlin2008602}. Along the way, a number of subshells (e.g. \emph{N}=40 in $^{68}$Ni \cite{BERNAS1982279} or \emph{N}=32, 34 in $^{52,54}$Ca \cite{Steppenbeck-Nature.502.207,Wienholtz-Nature.498.346}) have even been shown to exhibit localised magic-like behaviour.

Recent developments have revealed the importance of a more comprehensive approach to the study of nuclear structure. The example of $^{52}$Ca is telling : while the relatively large E(2$^{+}_{1}$) \cite{PhysRevC.74.021302} and large empirical shell gap \cite{Wienholtz-Nature.498.346} indicate doubly magic character, recent high-resolution measurements of the electromagnetic moments and mean-squared charge radius \cite{Rui16} do not. However, the question of declaring $^{52}$Ca ``magic" or not is less important than the rather peculiar combination of ground state properties which makes this nucleus a real challenge for theory. 
 
With hints of a shell closure in $^{68}$Ni and $^{52,54}$Ca and a variety of phenomena in between, this region between the magic proton numbers 20 and 28 is the perfect playground to benchmark rapidly developing \emph{ab initio} approaches, which have now reached \emph{Z}=50 \cite{Morris2017}. More significantly, this region marks the frontier of what is straightforwardly accessible to these methods \cite{Leistenschneider2017} and large-scale shell-model calculations \cite{Honma2005,PhysRevC.82.054301}. 

Unfortunately, the majority of observables currently available in the region is limited to excited states and transition probabilities, especially for the chromium isotopic chain which lies right in the middle. Similar to the \emph{N}=20 island of inversion, the properties of excited states along the \emph{N}=40 isotones suggest a rapid development of collectivity from a doubly magic $^{68}$Ni \cite{BERNAS1982279}, to a transitional $^{66}$Fe \cite{PhysRevC.81.061301} and finally a strongly deformed $^{64}$Cr \cite{PhysRevC.81.051304,PhysRevC.92.034306,PhysRevC.86.011305,PhysRevLett.110.242701}. Additionally, dominant collective behavior appears to persist past \emph{N}=40 \cite{PhysRevLett.115.192501} possibly merging the \emph{N}=40 island of inversion with a region of deformation in the vicinity of $^{78}$Ni \cite{PhysRevLett.117.272501}. 

Below $^{68}$Ni, high-precision mass spectrometry studies have reached down as far as the manganese chain \cite{PhysRevC.86.014325,PhysRevC.81.044318}, while laser-spectroscopy information is only available for the manganese isotopes \cite{PhysRevC.94.054321,PhysRevC.92.044311,BABCOCK2016387}. Time-of-flight mass measurement results in the atomic mass evaluation 2016 \citep{AME2016} suggest a sudden onset of deformation towards \emph{N}=40 in the chromium chain. However, the data are not precise enough to draw reliable nuclear-structure conclusions. 

The need for accurate and precise mass values of neutron-rich chromium isotopes is also of interest in the field of astrophysics. Neutron-rich chromium masses can play an important role in the cooling and heating of the crust of accreted neutron stars possibly impacting the associated astrophysical observables. Indeed, the electron-capture sequence $^{64}$Cr$\rightarrow^{64}$V$\rightarrow^{64}$Ti was shown to be the main heating source in the lower layers of the outer crust \cite{PhysRevC.93.035805,0004-637X-662-2-1188}. Additionally, nuclides in regions of strong deformation, such as $^{59}$Cr and $^{63}$Cr, were shown \cite{Schatz.nature} to allow for strong Urca cooling \cite{PhysRev.59.539} cycles.

In this Letter, we report the first precision measurements of the ground-state binding energies of short-lived neutron-rich chromium isotopes. Our results are compared to all major families of nuclear models, including density functional theory and large-scale shell model, based on the LNPS interaction as well as \emph{ab initio} calculations from the recently developed nucleus-dependent valence-space formulation of the in-medium similarity renormalization group (VS-IMSRG) \cite{Stro17ENO}.

The chromium isotopes measured in this work were produced at the ISOLDE/CERN \cite{ISOLDE_2017} radioactive ion beam facility as fission products from the interaction of a 1.4-GeV proton beam impinging on a thick UC$_x$ target. The target was heated to enable the species of interest to diffuse into a dedicated ionization region. For the first time at an ISOL facility, chromium ion beams were produced by a resonance ionization laser ion source (RILIS) \cite{1993550}. A three-step scheme was developed, the details of which can be found in \cite{DayGoodacre201758}. The chromium ions were transported to the ISOLTRAP setup \cite{Mukherjee-EurPhysJA,Kreim-NuclInstrumMethodsB.317.492}, at a kinetic energy of 30 keV, via the High-Resolution Separator (HRS) magnets of the ISOLDE facility. Ions were accumulated in a linear radio-frequency quadrupole cooler and buncher for 10 ms \cite{Herfurth2001254}. The resulting ion bunches were subsequently decelerated by a pulsed drift cavity to an energy of 3.2 keV before being injected into a multi-reflection time-of-flight mass separator (MR-ToF MS) \cite{Wolf-NuclInstrumMethodsA.686.82}. Inside the MR-ToF MS, the ion bunch underwent typically 1000 revolutions between two electrostatic mirrors separating the isobaric constituents of the bunch. In this work, the molecular contaminants CaF$^{+}$ and TiO$^{+}$ were predominant in the ISOLDE beam. To provide the subsequent Penning traps with purified beams the ejection timing of the MR-ToF MS was optimized to suppress contamination \cite{WIENHOLTZ2017285}. After cooling inside a preparation Penning trap \cite{SAVARD1991247}, the chromium ions were delivered to ISOLTRAP's precision Penning trap where the mass measurements were carried out. 

\begin{figure}[!h]
\centering
\includegraphics[scale=0.45]{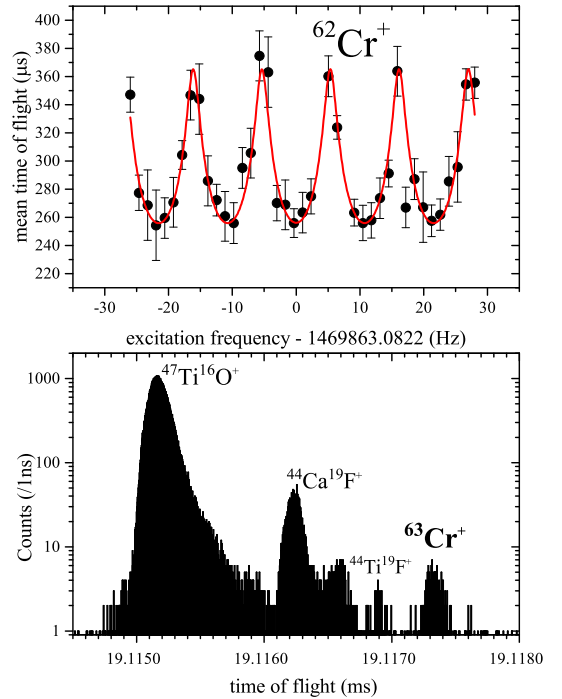}
\caption{(Color online){\it Top}: A typical ToF-ICR resonance of $^{62}$Cr$^{+}$ ions 
using a Ramsey-type excitation scheme (10 ms-80 ms-10 ms \cite{George-IntJMassSpectrom.264.110}). The line 
represents a fit to the data points.
{\it Bottom}: Number of events as a function of flight time after 1000 revolutions of the $A=$63 ISOLDE beam inside the MR-ToF MS.}
\label{spectra}
\end{figure}

\begin{table*}[t]
\centering
\resizebox{1\textwidth}{!}{\begin{minipage}{\textwidth}
 \caption{Frequency ratios ($r = \nu_{c,ref} / \nu_{c}$), time-of-flight ratios ($C_{ToF}$) and mass excesses of the chromium isotopes measured in this work. Values of the mass excesses from the Atomic-Mass Evaluation 2016 (AME2016) \cite{AME2016} are given for comparison.
The masses of the reference ions were also taken from AME2016. Experimental half-lives are from \cite{Nubase2016}. The yields correspond to the number of chromium ions per second delivered by the ISOLDE facility. The total transport efficiency of the experiment was estimated to be 0.5\% behind the MR-ToF MS.The average proton current on target was 1.8$\mu$A.}
\begin{ruledtabular}
  \begin{center}
    \begin{tabular}[c]{l c c c c c c}
          &                     &                 &                 &                            & \multicolumn{2}{c}{\underline{Mass Excess (keV)}}        \\ 
     Ion    & Yield (Ions/s) & Half-life  & Reference      & Ratio $r$ or $C_{ToF}$     & New               & AME16                         \\ \hline\hline
        $^{58}$Cr\footnote{Mass value included in the 2016 mass evaluation as a private communication.}   & Not determined & 7.0(0.3) s & $^{85}$Rb  & r = 0.6824024142(376) & -51 991.8(3.0) & -51 991.8(1.5) \\ \hline
     \multirow{2}{*}{$^{59}$Cr}  & \multirow{2}{*}{$3 \times 10^{5}$} & \multirow{2}{*}{1050(90) ms}  & $^{85}$Rb   & r = 0.6942284208(85)  & -48 115.9(0.7) & \multirow{2}{*}{-48 090(220)} \\ 
& & & $^{40}$Ca$^{19}$F/$^{85}$Rb   & $C_{ToF}$ = 0.500536923(887)  & -48 132(20) & \\ \hline
	 \multirow{2}{*}{$^{60}$Cr}  & \multirow{2}{*}{$2 \times 10 ^{4}$} & \multirow{2}{*}{490(10) ms}  & $^{85}$Rb   & r = 0.7060206906(138)  & -46 908.5(1.1) & \multirow{2}{*}{-46 670(190)} \\
& & & $^{41}$Ca$^{19}$F/$^{85}$Rb   &$C_{ToF}$ = 0.500484920(886)  & -46 917(19)  &  \\ \hline
	 \multirow{2}{*}{$^{61}$Cr}  & \multirow{2}{*}{$2 \times 10^{3}$} & \multirow{2}{*}{243(9) ms}  & $^{85}$Rb  & r = 0.7178534753(230) & -42 496.5(1.8) & \multirow{2}{*}{-42 480(100)} \\
& & & $^{42}$Ca$^{19}$F/$^{85}$Rb   &$C_{ToF}$ = 0.500120578(956)  & -42 503(20)  &  \\ \hline
	\multirow{2}{*}{$^{62}$Cr}  & \multirow{2}{*}{$3 \times 10^{2}$} & \multirow{2}{*}{206(12) ms}  & $^{85}$Rb  & r = 0.7296512630(440) & -40 852.6(3.5) & \multirow{2}{*}{-40 890(150)} \\
& & & $^{43}$Ca$^{19}$F/$^{85}$Rb   &$C_{ToF}$ = 0.500047948(922)  & -40 841(18) &  \\ \hline
     $^{63}$Cr  & $3 \times 10 ^{1}$              & 129(2) ms        & $^{44}$Ca$^{19}$F/$^{85}$Rb  & $C_{ToF}$ = 0.49964187(386) & -36 178(73)      & -36 010(360) \\
    \end{tabular}
   \end{center}
   \label{ResTable}
  \end{ruledtabular}
  \end{minipage}}
\end{table*}

The mass determination relies on the measurement of the cyclotron frequency 
$\nu_c = q{ B}/(2\pi m)$ of an ion with mass
$m$ and charge $q$ in a magnetic field of strength $B$. Before and after each measurement of chromium ions, the cyclotron frequency of a reference ion was determined. In this experiment, $^{85}$Rb$^{+}$ ions provided by ISOLTRAP's offline ion source were used. The masses of $^{58-62}$Cr were measured using the ToF-ICR technique, both in the one-pulse excitation \citep{Koenig-IntJMassSpectrom.142.95} and two-pulse Ramsey excitation scheme \cite{George-IntJMassSpectrom.264.110} (upper panel of Figure \ref{spectra}).

In the case of $^{63}$Cr, the production yield was so low that the mass determination could only be performed using ISOLTRAP's MR-ToF MS as a mass spectrometer. The ions' time of flight was recorded with an electron multiplier placed behind the MR-ToF MS. The relationship between a singly charged ion's flight time $t$ and its mass $m$ is given by $t = a\times(m)^{1/2} + b$. The parameters $a$ and $b$ can be determined by measuring the flight times $t_{1,2}$ of reference ions with well-known masses $m_{1,2}$ following the same number of laps in the MR-ToF MS as the ion of interest.
The mass of the ion of interest is then determined as \cite{Wienholtz-Nature.498.346} : $ \sqrt{m} = C_{ToF}\times(\sqrt{m_1} -\sqrt{m_2}) + 0.5\times(\sqrt{m_1} + \sqrt{m_2})$, where $C_{ToF} = (2t - t_1 -t_2)/[2(t_1 - t_2)]$. In the present work the masses of $^{59-63}$Cr were determined in this fashion. Each time-of-flight spectrum was calibrated using the isobaric CaF$^{+}$ ions and offline $^{85}$Rb$^{+}$ ions which were measured after the same number of revolutions inside the MR-ToF device. An example of a time-of-flight spectrum is presented in the bottom panel of Figure \ref{spectra}. 

Due to the low statistics encountered in some spectra the analysis was performed in all cases using the binned maximum likelihood method assuming a Gaussian peak shape. In total $\sim$ 2000 $^{63}$Cr ions were recorded over 7 spectra, some containing as few as 90 counts. Typically 1000 revolutions inside the MR-ToF MS were used but with some variations (900-1200). The chromium-ion peak was identified unambiguously by blocking the first step of the RILIS scheme. Isobaric references were identified by the measurement of their cyclotron frequencies in the precision Penning trap. In all cases, the contribution associated with the offline species to the statistical uncertainty was evaluated to be more than three orders of magnitude smaller than that associated with the online species. As a result, effects related to relative time-of-flight fluctuations between offline and online species are by far below the current statistical precision of the measurements. Nonetheless, to keep these effects under control, an offline reference spectrum was measured shortly before the acquisition of every single online spectrum. The asymmetries of the peak were taken into account on a case-by-case basis by evaluating the influence of the range on the fit results. Systematic contributions to the $C_{ToF}$ uncertainty were evaluated from on-line data to be 0.87$\times 10 ^{-6}$. An additional systematic uncertainty accounting for possible time-of-flight drifts arising from coulomb interaction between reference species and ions of interest was found to only play a significant role for $^{63}$Cr, for which an additional $C_{ToF}$ uncertainty of 3.74$\times 10 ^{-6}$ was added in quadrature. The final frequency and/or time-of-flight ratios, as well as their associated uncertainties corresponding to one standard deviation, are listed in Table \ref{ResTable}. The mass excesses of $^{59-62}$Cr determined with the MR-ToF MS agree within their uncertainties with the values from the well-established ToF-ICR technique. 

\begin{figure}
\centering
\includegraphics[scale=0.45]{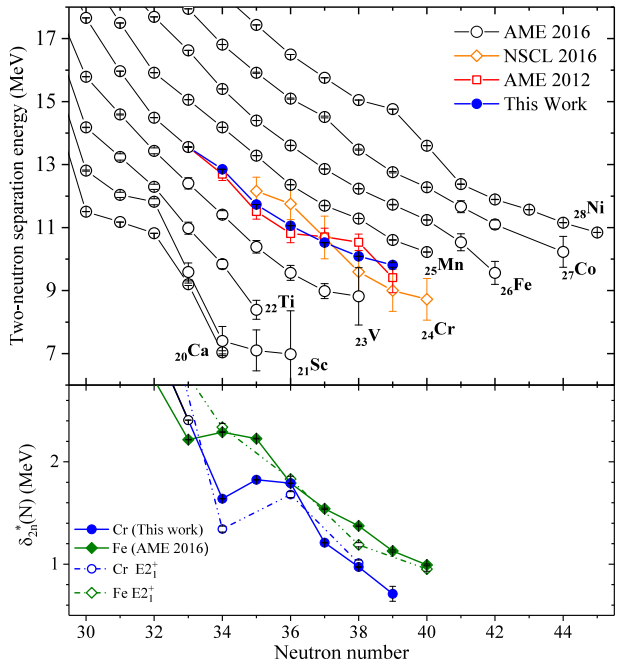}
\caption{(Color online){\it Top:} The $S_{2n}$ trend for Cr from AME2016 \cite{AME2016} is decomposed between the results compiled in the 2012 mass evaluation \cite{AME2012} and recent results from the S800 spectrometer at the NSCL \cite{PhysRevC.93.035805}.
{\it Bottom:} The trend of $\delta_{2n}^{*}(N,Z)$ for the ground and 2$_{1}^{+}$ excited states \cite{PhysRevC.81.051304,PhysRevLett.115.192501} in the Cr and Fe chains.}
\label{S2n_fit}
\end{figure}

\begin{figure}
\includegraphics[scale=0.45]{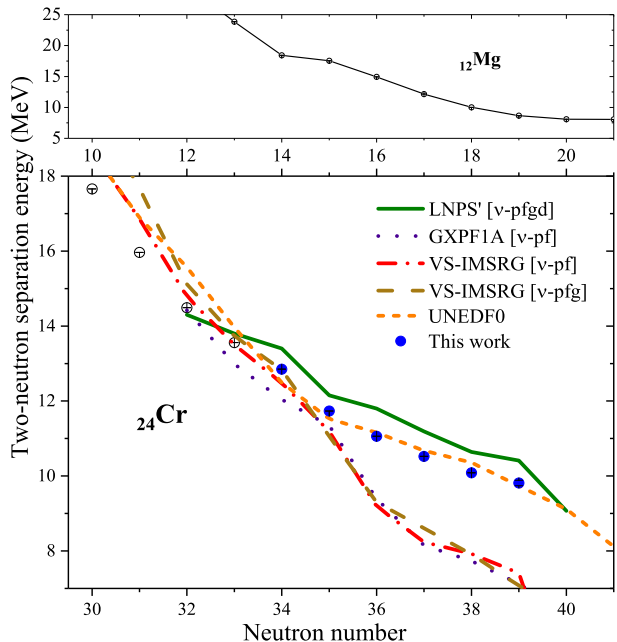}
\caption{(Color online){\it Top:} Two neutron-separation energies of the Mg chain \cite{AME2016}.{\it Bottom:} Comparison between the experimental $S_{2n}$ values for the chromium isotopes and predictions from various nuclear models. For shell-model Hamiltonians, the corresponding neutron valence space is indicated in brackets.}
\label{S2n_sm}
\end{figure}

Figure \ref{S2n_fit} shows the two-neutron separation energies, defined as $S_{2n}(N,Z) = ME(N-2,Z)-ME(N,Z)+2M_{n}$, for the region of interest where $ME(N,Z)$ represents the mass excess of an isotope with \textit{N} neutrons and \textit{Z} protons and $M_{n}$ is the neutron mass excess. The AME2012 \cite{AME2012} suggests the presence of a kink at \textit{N}=38 in the $S_{2n}$ trend. The trend obtained in this work shows a markedly different behaviour. It appears very smooth with an upward curvature when approaching \emph{N}=40, suggesting that an onset of collectivity is visible in the ground state of neutron-rich Cr isotopes. The trend of $S_{2n}$ observed closely resembles that of Mg in the original island of inversion from \emph{N}=14 to 20, shown in the upper panel of Figure \ref{S2n_sm}.

To quantify the $S_{2n}$ trend, the lower panel of Fig. \ref{S2n_fit} shows the quantity $\delta_{2n}^{*}(N,Z) = S_{2n}(N-2,Z) – S_{2n}(N,Z)$. A low value of $\delta _{2n}^{*}$ marks a flattening of $S_{2n}$ so the reduction of $\delta_{2n}^{*}$ is usually associated with an increase of collectivity by a gain in correlation energy \cite{PhysRevC.82.061306}. The $\delta_{2n}^{*}$ is plotted both for the ground and 2$^{+}$ states of the Fe and Cr isotopes. One notices that for \emph{N} $>$ 36 the $\delta_{2n}^{*}$ values of Cr (both ground and 2$^{+}$ state) are smaller than the ones of Fe isotopes, meaning a stronger flattening of $S_{2n}$ and a faster correlation-energy gain. The values of ground and 2$^{+}$ states for each isotopic chain are relatively close, but at the onset of collectivity observed in the E2$^{+}_{1}$trends (\emph{N} = 36 for Cr and \emph{N} = 38 for Fe \cite{PhysRevC.81.051304}), the ground-state  $\delta_{2n}^{*}$ is higher, illustrating a slower gain in correlation energy.

$S_{2n}$ trends obtained from different theoretical approaches are presented in the lower panel of Figure \ref{S2n_sm}.
Mean-field calculations of even-even and odd-even chromium isotopes were performed using the UNDEF0 energy-density functional \cite{PhysRevC.82.024313}. The HFBTHO code, which solves the HFB equations enforcing axial symmetry \cite{STOITSOV20131592}, was used. The odd-\emph{N} isotopes were computed performing quasi-particle blocking within the so-called equal-filling approximation \cite{PhysRevC.78.014304}. The Lipkin-Nogami prescription was used for approximate particle-number restoration. The UNEDF0 predictions are in very good quantitative agreement for $N$ $>$ 33. Additionally, axial deformation quadrupole moment-constrained HFB calculations were performed for the even-\emph{N} chromium isotopes. Our calculations predict a spherical ground-state for  $^{64}$Cr and show the development of static deformation starting only at $^{68}$Cr ($\beta \approx$ 0.2). This is consistent with previous predictions using the Gogny D1S functional \cite{PhysRevC.80.064313,PhysRevC.93.054316} where the importance of beyond-mean-field approaches to adequately describe collectivity in the chromium chain was highlighted.

The traditional approach of the nuclear configuration interaction treats the dynamics of valence particles outside an inert core using a phenomenological Hamiltonian optimized to a specific valence space. This approach can access many of observables and is applicable far from closed shells, provided the diagonalization remains computationally tractable. In the phenomenological shell model, the interplay between the central field and the tensor component of the nuclear interaction was proposed to be responsible for the shell-evolution phenomenon \cite{PhysRevLett.87.082502,PhysRevLett.95.232502,Smirnova2010109}. A reduction of the associated spherical shell gaps may give rise to a deformed ground state provided that correlations are sufficient to energetically favor an intruder state. The marked discrepancy between the $S_{2n}$ trend obtained in this work and the one predicted using the GXPF1A phenomenological interaction \cite{Honma2005} highlights the need to include the $d_{5/2}$ and $g_{9/2}$ orbitals in the valence space in order to reach agreement between theory and experiment. The LNPS interaction \cite{PhysRevC.82.054301} was derived specifically to correct for this deficiency. The valence space adopted is therefore based on a $^{48}$Ca core and includes the full \emph{pf} shell for protons and the neutron $pf_{5/2}g_{9/2}d_{5/2}$ orbitals \cite{PhysRevC.82.054301}. LNPS$^{\prime}$ is a version in which the global monopole term was made 30 keV more attractive in order to reproduce the $S_{2n}$ trend obtained in Ref.~\cite{PhysRevC.93.035805}. Nonetheless, the LNPS$^{\prime}$ trend shows a systematic over-estimation of the $S_{2n}$ values of almost 1 MeV compared to the present work.

We have also performed \emph{ab initio} calculations using the VS-IMSRG  \cite{Tsuk12SM,Bogn14SM,Stro16TNO,Stro17ENO,Simo17SatFinNuc}, which allow us to test nuclear forces in fully open-shell systems. With four protons outside the nearest closed shell the chromium isotopes are currently beyond the reach of other large-space \emph{ab initio} methods. We start from the 1.8/2.0 (EM) NN+3N interactions developed in Refs.~\cite{Hebe11fits,Simo17SatFinNuc}, which reproduce well ground-state properties of nuclei throughout the sd and pf regions \cite{Simo16unc,Hag16,Rui16,Simo17SatFinNuc}. Unless specified, details of the calculations are the same as those given in Ref.~\cite{Simo17SatFinNuc}. We then use the Magnus formulation of the IMSRG \cite{Morr15Magnus,Herg16PR} to decouple the $^{40}$Ca core as well as a $pf$-shell valence space Hamiltonian. With the ensemble-normal ordering introduced in Ref.~\cite{Stro17ENO}, we include effects of 3N forces between valence nucleons, such that a different valence space Hamiltonian is constructed specifically for each nucleus. The final diagonalization was performed using the NuShellX@MSU shell-model code \cite{BROWN2014115}. The lower panel of Figure \ref{S2n_sm} shows the $S_{2n}$ derived within this framework. 

A marked discrepancy between the theoretical and experimental trends obtained in this work is observed starting at $\textit{N} \geqslant 35$. It is worth noting that the VS-IMSRG trend closely resembles that obtained with the phenomenological GXPF1A interaction, hinting that the VS-IMSRG interaction is missing the same degrees of freedom for describing the physics in the region. To investigate this possibility, we have extended the VS-IMSRG for the first time to allow for mixed-parity valence spaces.  As in the LNPS interaction, by excluding the neutron $f_{7/2}$  orbital, we are then able to include the neutron $g_{9/2}$ orbital in the valence space. Unfortunately, we see that the addition of the $g_{9/2}$ orbital does not significantly improve the calculations. An attempt to include the neutron $d_{5/2}$ orbital was made. However, the dramatic increase in dimension is such that above $^{60}$Cr the calculation becomes intractable. Nonetheless, the calculation for $^{60}$Cr shows that the additional neutron $d_{5/2}$  orbital is occupied at $\sim$2\%, a value similar to the one obtained for the LNPS interaction and in Gogny D1S calculations \cite{PhysRevC.83.054312}. The improvement of $S_{2n}$ was only of around 100 keV, insufficient to correct the discrepancy observed in Fig. \ref{S2n_sm}. Consequently, the absence of the $d_{5/2}$  orbital cannot alone be responsible for the observed discrepancy at and beyond \emph{A}=60. Difficulty in capturing deformation in the VS-IMSRG calculations is only now being explored, but likely originates in neglected three- or higher-body operators, the physics of which is not built into the valence-space Hamiltonian \cite{Henderson18}. Work in this direction is currently in progress.
 
In conclusion, benefiting from the laser-ionized chromium ion beam developed at the ISOLDE facility for this experiment, the masses of the neutron-rich isotopes $^{58-63}$Cr were measured using precision mass-spectrometry techniques. The present mass values are a factor of up to 300 times more precise than the ones in the literature thus greatly refining our knowledge of the mass surface in the vicinity of \textit{N}=40. We exclude a sudden onset of deformation showing rather a gradual enhancement of ground-state collectivity in the chromium chain. We illustrate that, in the region of enhanced collectivity, the gain in binding per added neutron is more significant in the Cr chain than in the Fe chain and slightly more pronounced in the excited 2$^{+}$ states. The evolution of the $S_{2n}$ trend is well reproduced by both the UNEDF0 density functional and the LNPS$^{\prime}$ phenomenological shell model interaction. For the first time, we apply the \emph{ab initio} VS-IMSRG to open-shell chromium isotopes, exploring extensions to mixed-parity valence spaces. The precise data thus provide important constraints to guide the ongoing development of \emph{ab initio} approaches to nuclear structure. 

\begin{acknowledgments}
We thank K. Sieja for stimulating discussions. MM thanks A.Pastore for his help with the mean-field calculations presented in this work. We thank the ISOLDE technical group and the ISOLDE Collaboration for their support. We acknowledge support by the French IN2P3, the BMBF (05P12HGCI1, 05P12HGFNE, 05P15HGCIA, 05P15RDFN1), the ERC Grant No.~307986 STRONGINT, the Max-Planck Society, the NRC and NSERC of Canada. Computations were performed with an allocation of computing resources at the J\"ulich Supercomputing Center (JURECA) and at the Max-Planck-Institute for Nuclear Physics. This work was supported in part by the DFG through the Cluster of Excellence Precision Physics, Fundamental Interactions and Structure of Matter (PRISMA). This project has received funding through the European Union’s Seventh Framework Programme for Research and Technological Development under Grant Agreements: 267194 (COFUND), and 289191 (LA3NET). This project has received funding from the European Union’s Horizon 2020 research and innovation programme under grant agreement No 654002. This project has also received funding from the European Union’s 7th framework through ENSAR under grant agreement No262010. 
\end{acknowledgments}

\bibliography{59-63Cr}

\end{document}